\documentclass{elsart}
\usepackage{amsmath,amssymb,graphics,epsfig}

\newcommand{\sign}[0]{\text{sign}}
\newcommand{\bra}[1]{\left\langle #1\right |}
\newcommand{\ket}[1]{\left| #1\right \rangle}
\newcommand{\braket}[2]{\left\langle #1 | #2 \right\rangle}

\begin{document}
\begin{frontmatter}
\title{Low-lying Wilson Dirac operator eigenvector mixing in dynamical overlap Hybrid Monte-Carlo}
\author{Nigel Cundy}\footnote{Email: nigel.cundy@physik.uni-regensburg.de\\Tel: +49 (0)202 9432014}
\address{Institut f\"ur Theoretische Physik,Universit\"at Regensburg,D-93040 Regensburg, Germany}

\begin{abstract}
Current dynamical overlap fermion hybrid Monte Carlo simulations encounter large fermionic forces when there is mixing between eigenvectors of the kernel operator with near zero-eigenvalues. This leads to low acceptance rates when there is a large density of near zero eigenvalues. I present a method where these large  forces are eliminated and the large action jumps seen when two eigenvalues approach zero are significantly reduced. This significantly increases the stability of the algorithm, and allows the use of larger integration time steps.
\end{abstract}
\end{frontmatter}

\newpage
\section{Introduction}\label{sec:1}
The overlap Dirac operator\cite{Narayanan:1993ss,Neuberger:1998fp}, which unlike other formulations of lattice QCD has an exact lattice chiral symmetry~\cite{Luscher:1998pqa} and a corresponding index theorem, offers numerous exciting possibilities for research in dynamical lattice QCD~\cite{Fodor:2003bh,Cundy:2005pi,DeGrand:2004nq,Hashimoto:2006rb}; but presents a number of distinct challenges. The first challenge is the numerical cost, but this is not insurmountable on modern computers. Today simulations on $16^332$ lattices are feasible~\cite{Fukaya:2007fb}, and it will not be long until large scale simulations will not only be possible but entirely practical and commonplace. The other difficulties involve the technical details of the algorithm, and in this paper I will focus on one of these issues, so far unexplored in the literature.

The overlap operator is defined as
\begin{gather}
D = (1+\mu) + (1-\mu)\gamma_5\epsilon(Q),
\end{gather}
where $\mu$ is a mass parameter proportional to the bare fermion mass and $Q$ is the Hermitian form of a suitable lattice Dirac operator (the kernel) with no fermion doublers and negative mass $\rho$. In this work, I will always use the Wilson operator with $\rho = 1.5,$ or, alternatively, $\kappa = 1/(8-2\rho) = 0.2$:
\begin{gather}
Q_{xy} = \gamma_5 \left[\delta_{xy}-\kappa\sum_{\mu}\left( (1-\gamma_{\mu})U_{\mu}(x)\delta_{y,x+\mu} + (1+\gamma_{\mu})U^{\dagger}_{\mu}(x-\mu)\delta_{y,x-\mu}\right)\right].
\end{gather}

  The matrix sign function is defined as
\begin{gather}
\epsilon(Q) = \sum_{i}|\psi_i\rangle\langle \psi_i|\text{sign}(\lambda_i),
\end{gather}
where $|\psi_i\rangle$ and $\lambda_i$ are the eigenvectors and eigenvalues of $Q$ respectively, and the sum is over the complete set of eigenvectors. In practice, given that the calculation of the entire eigenvalue spectrum is impractical, it is usual to use an approximation to the sign function, such as the Zolotarev Rational approximation ~\cite{vandenEshof:2002ms}, for the bulk of the eigenvalue spectrum. The spectral decomposition is only used for for the eigenvalues closest to zero, where no approximation can (realistically) be accurate enough without a large computational cost.
 
In terms of the Hermitian overlap operator $H = \gamma_5 D$, and the gauge action $S_g[U]$ for a gauge field $U$, the lattice QCD partition function for two degenerate flavours of fermion is
\begin{gather}
Z = \int d U \det (H^2[U,\mu])e^{-S_g[U]} = \int dU d\phi d\phi^{\dagger}e^{-S_g[U]-\phi^{\dagger}H^{-2}\phi},
\end{gather}
where I have used pseudo-fermion fields $\phi$ to approximate the fermion determinant. The standard Hybrid Monte Carlo (HMC) algorithm~\cite{HMC} generates a new gauge field by introducing a momentum $\Pi$, updating the momentum and gauge field along the classical trajectory using a numerical integration algorithm (the molecular dynamics), and finishing with a metropolis step to ensure that the update of the gauge field satisfies detailed balance. The numerical integration must be reversible and ergodic. It does not have to be area conserving, but in a non-area conserving molecular dynamics the Jacobian must be calculated and included in the metropolis accept/reject step, as discussed in section ~\ref{sec:2}.

The numerical integration requires the calculation of a fermionic force, obtained by differentiating the action with respect to the gauge field. For the overlap operator, the action is discontinuous, leading to two problems: firstly there is a delta function in the force whenever an eigenvalue of the kernel operator, $Q$, changes sign; and secondly a large peak in the force when two eigenvectors, whose eigenvalues have different signs, mix. The first problem can be compensated for using the ``transmission/reflection" algorithm, first published by Zoltan Fodor and collaborators~\cite{Fodor:2003bh}, and subsequently improved by my own work~\cite{Cundy:2005pi,Cundy:2005mr}. There are still additional difficulties, particularly the rate of topological charge changes at small mass~\cite{Egri:2005cx} and the volume dependence of the algorithm~\cite{Schaefer:2006bk}, but these can be resolved~\cite{cundyforthcoming,Cundy:2007la}. 

\begin{figure}
\begin{center}
\begin{tabular}{c}
\includegraphics[height = 6.0cm]{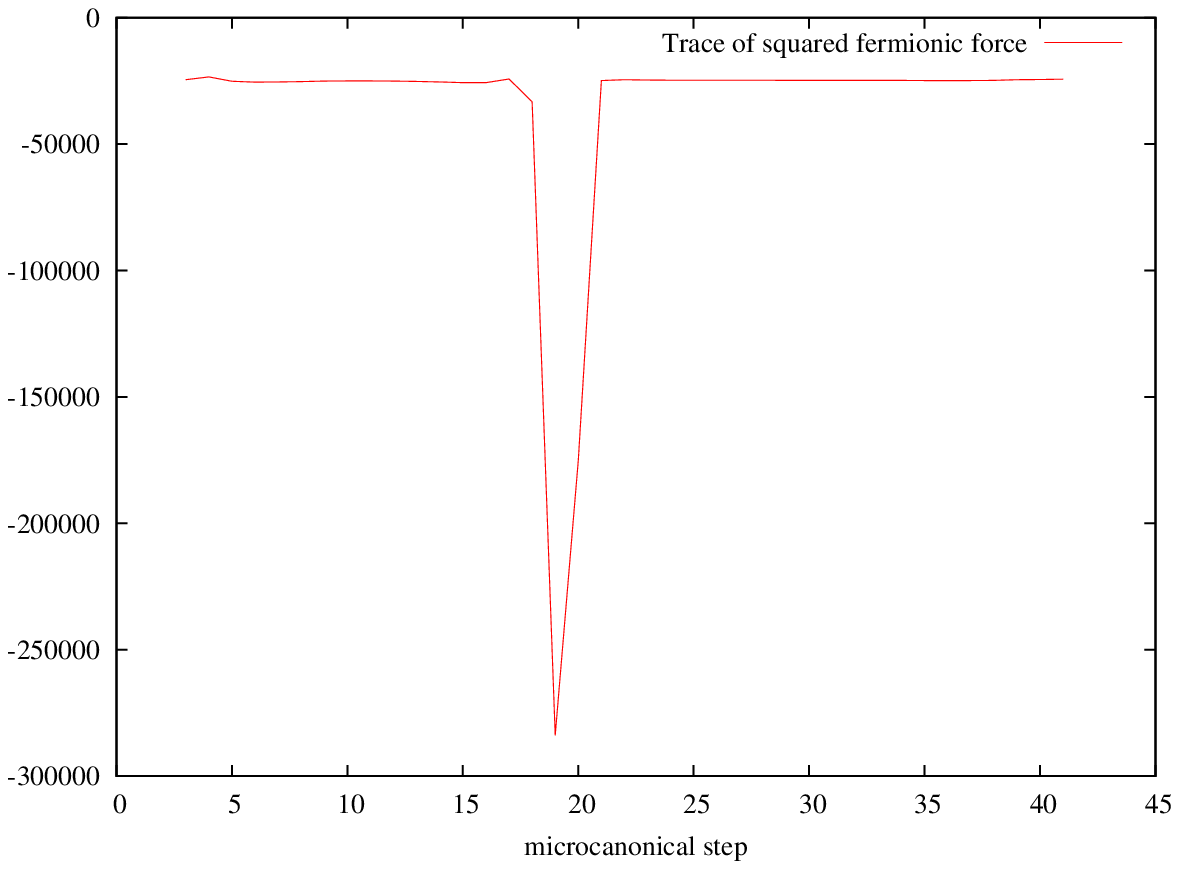}\\
\includegraphics[height = 6.0cm]{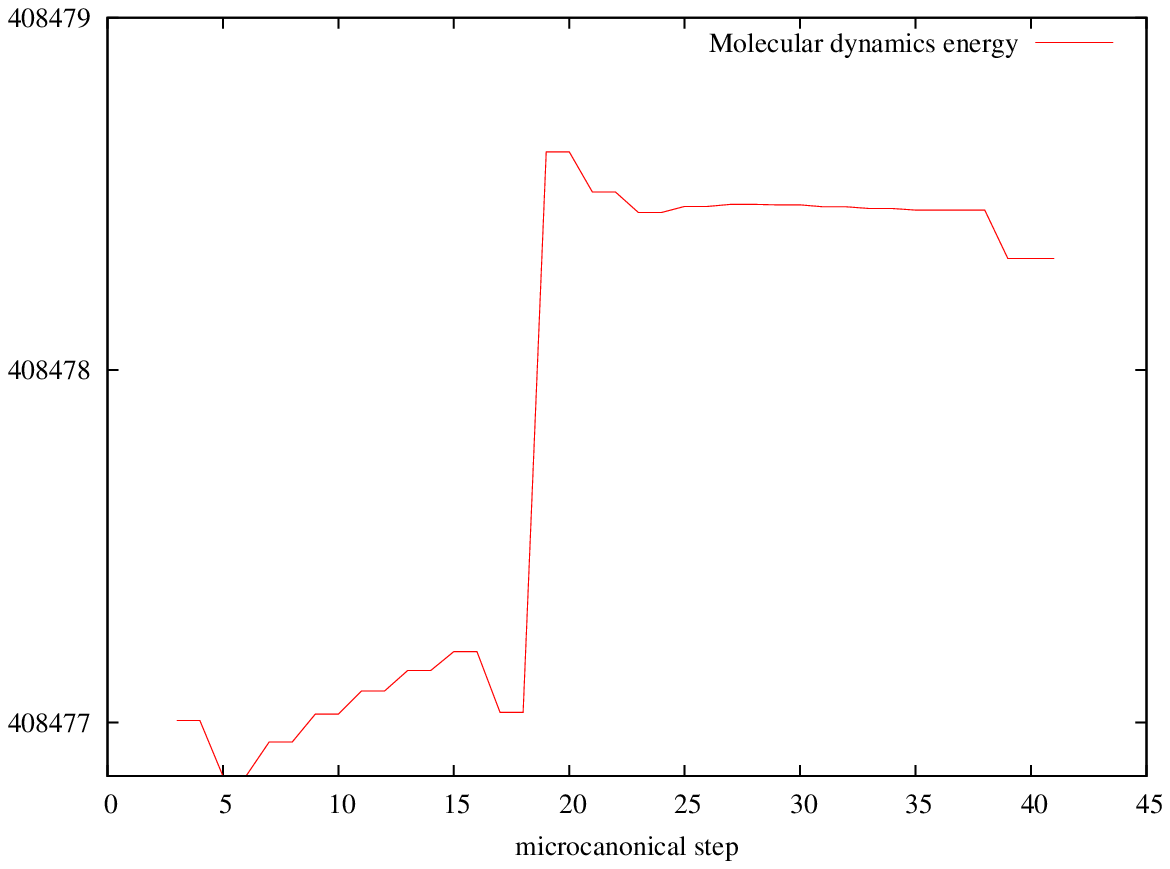}\\
\includegraphics[height = 6.0cm]{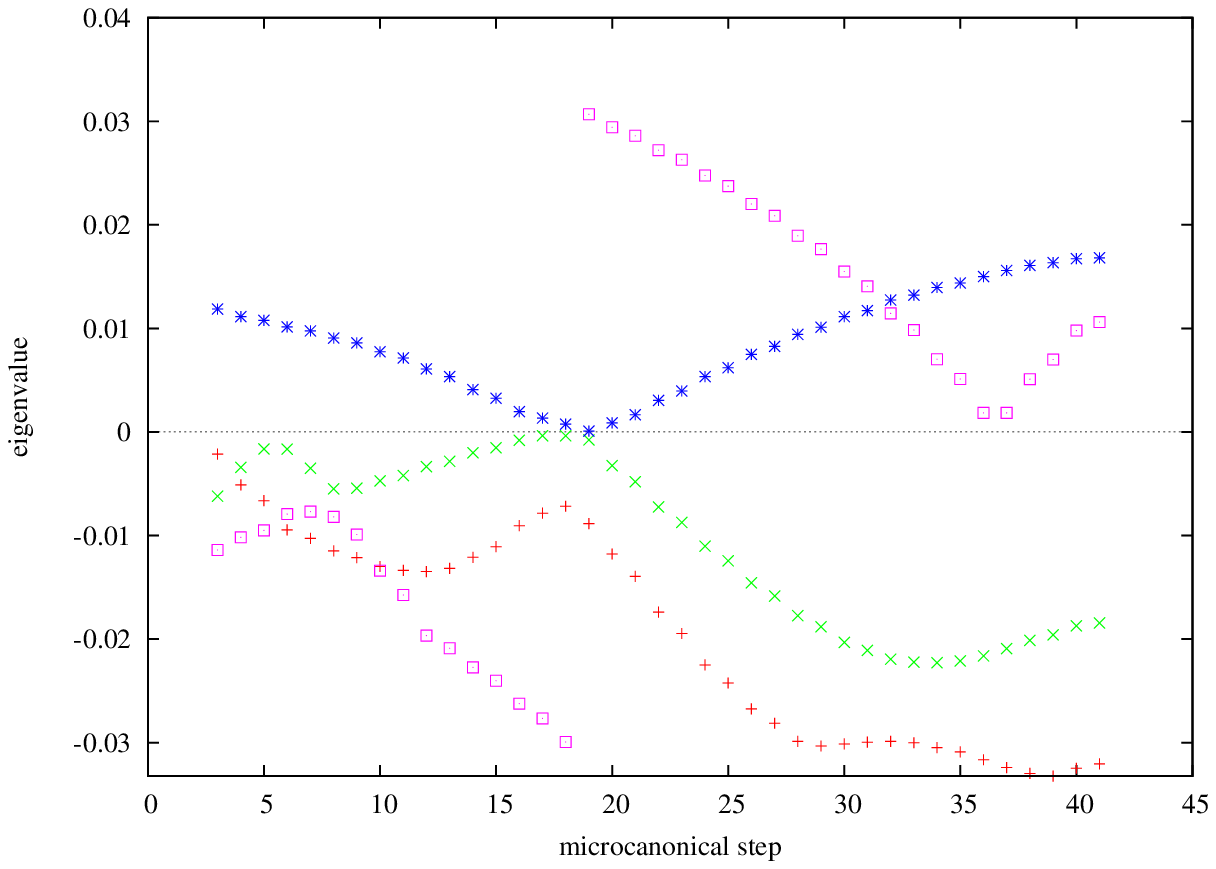}
\end{tabular}
\end{center}
\caption{The trace of the square of the fermionic force (top), molecular dynamics energy (middle), and the Wilson operator eigenvalues (bottom) across one trajectory on an $8^316$ ensemble with mass $\mu = 0.03$, time-step $\tau = 0.01$, two pseudo-fermion fields, and two steps of stout smearing with parameter 0.1. By explicit calculation during the molecular dynamics, I observed that there was no exactly zero Wilson eigenvalue between between the 7th and 35th micro-canonical steps. The two low lying eigenvectors mixed at the 19th micro-canonical step, but the eigenvalues did not cross.}\label{fig:oldforce}
\end{figure}

The second problem is a little more technical. Until this study, the eigenvalues and eigenvectors of a sparse matrix have been differentiated using a procedure analogous to first order perturbation theory. This method is outlined in reference~\cite{Cundy:2005mr}, although the idea is not original to the cited paper.  The differential of the matrix sign function (neglecting the delta function) with respect to the molecular dynamics time $\tau$ obtained from this method can be expressed in terms of the complete basis of eigenvalues and eigenvectors of the kernel operator
\begin{gather}
\frac{d}{d\tau}\left( |\psi_i\rangle\langle \psi_i|\right) \sign(\lambda_i)= \sum_{j\neq i}|\psi_j\rangle\langle \psi_j|\frac{d}{d\tau} Q |\psi_i\rangle\langle \psi_i|\frac{\sign(\lambda_i) - \sign(\lambda_j)}{(\lambda_i - \lambda_j)}.\label{eq:1}
\end{gather}
It is clear that there is a large differential, and thus large fermionic force, when there is a pair of eigenvalues close to zero, but with different signs (see figure \ref{fig:oldforce}). I refer to this as the ``eigenvalue mixing problem" for reasons that shall become obvious later. So far dynamical overlap simulations have tried to avoid this problem by suppressing the number of small eigenvalues of the kernel Dirac operator, either by smearing~\cite{DeGrand:2004nq,Morningstar:2003gk}, or by adding an additional term to the action ~\cite{Fukaya:2006vs}.\footnote{These two approaches also have the advantage of accelerating the computation.} Neither of these methods are satisfactory: too much smearing will distort the physics, and will not remove the problem on sufficiently large volumes. The algorithm with the additional term may not be ergodic if topological sectors are either not internally connected, or (if they are connected) the computer time required to evolve to a different sub-sector is unreasonably large. My own approach so far has been to use a moderate amount of smearing, regulate the force to prevent it from becoming too large (leading to instabilities and a breakdown of reversibility), and to run short trajectories (so that if I do encounter a problem I have not lost too much computer time) with a small time-step (which, as shall be made clear later, reduces the number of occurrences of the large forces). This allowed me to run on small lattices (up to $12^324$), with the large forces sufficiently infrequent that they did not significantly reduce the metropolis acceptance rate. However, as the lattice volume is increased, the density of small eigenvalues also increases and the time-step would have to be reduced to unmanageable proportions to allow acceptance. Also, methods which use multiple times scales~\cite{SW} combined methods such as using additional pseudo-fermions to precondition the force~\cite{Hasenbusch:2001ne} and RHMC~\cite{Clark:2004cp} are not as efficient as one might hope for. This is because the time-step needed for the integration is determined by the differential of the sign function, common to all the terms in the terms in the forces constructed in these methods, rather than the condition number of the overlap operator. Clearly reducing the time-step as the density of small eigenvalues increases is not an optimal solution.

The reason for these large forces becomes evident once it is realised that equation (\ref{eq:1}) is just the first term in a Taylor expansion in $\tau/(\lambda_i - \lambda_j)$ of the mixing angle between the two eigenvectors, which is a function of the gauge field, time-step and momenta. Including higher order terms would lead to a force that does not conserve area or is not reversible. When $\tau/(\lambda_i - \lambda_j)$ is small, the expansion is valid, and everything works well. When it is not so small, the higher order terms start contributing, leading to an uncalculated and perhaps substantial correction to the energy conservation. When it is larger still, the series expansion may not converge at all. However, using the exact mixing angles rather than the expansion would eliminate the large forces.  In this paper, I describe how this can be done. This approach is not area conserving; but the Jacobian can be calculated, and corrected for in the metropolis accept/reject step. No account of the Jacobian is made when trying to conserve energy, but the size of the Jacobian contribution to the action is O($\tau^3$), the same as the normal molecular dynamics energy violations. When the mixing becomes large, there will be a large Jacobian, but this is still considerably smaller than the action jump caused by the large forces using the old method. This new method is not manifestly reversible, but it is possible to construct a reversible algorithm by combining forward and backward updates. Stout smearing is technically more challenging to apply efficiently with this new method; but it is possible.

Section \ref{sec:2} outlines how a non-area conserving (NAC) HMC can be constructed, and describes the calculation of the new fermionic force and the Jacobian. Section \ref{sec:3}  outlines numerical results comparing this algorithm with the old method. Section \ref{sec:4} is a conclusion, and there are two appendices describing some of the more technical details of the proposed algorithm.

\section{Non-area conserving HMC for overlap fermions} \label{sec:2}
\subsection{Hybrid Monte Carlo}
To fix the notation I start by reviewing the hybrid Monte Carlo algorithm for two flavours of fermion~\cite{HMC}. A Monte Carlo method satisfies the detailed balance condition
\begin{gather}
P[U'\leftarrow U] W_c[U] = P[U\leftarrow U'] W_c[U'],\label{eq:db}
\end{gather}
where $W_c[U]$ is the canonical ensemble and $P[U'\leftarrow U]$ is the probability of updating from gauge field $U$ to gauge field $U'$. In a Hybrid Monte Carlo method, we introduce a momentum field $\Pi$, which contains a Hermitian traceless matrix on every link of the lattice, and which is generated according to a Gaussian distribution. We evolve the gauge field and the momentum according to a reversible and ergodic trajectory $T[U,\Pi]$. Finally, we include a metropolis step to correct for small changes in the energy $E=\Pi^2/2 + S_g[U] + \phi^{\dagger}H^{-2}\phi$. Thus the probability of generating a field $U'$ from a field $U$, for a canonical ensemble,
\begin{gather}
W_c[U] = \int d\phi^{\dagger}d\phi e^{-S_g[U] - \phi^{\dagger}H^{-2}[U]\phi},
\end{gather}
is
\begin{align}
P[U'\leftarrow U]  =& \int d\Pi d\Pi' e^{-\frac{1}{2}\Pi^2}\delta([U',\Pi'] - T[U,\Pi]) \nonumber\\&\min\left(1,e^{-S_g[U'] - \phi^{\dagger}H^{-2}[U']\phi - \frac{1}{2}\Pi'^2 + S_g[U] + \phi^{\dagger}H^{-2}[U]\phi + \frac{1}{2}\Pi^2 - \log J}\right),\label{eq:prob}
\end{align}
where the fermion determinant is approximated using a pseudo-fermion field $\phi$, in the standard way, and $J$ is the Jacobian
\begin{gather}
J = \left|\begin{array}{l l}
\frac{\partial U}{\partial U'}&\frac{\partial\Pi}{\partial U'}\\ 
\frac{\partial U}{\partial \Pi'}&\frac{\partial\Pi}{\partial\Pi'}
\end{array}\right|.
\end{gather}
It is easy to show that this update satisfies the detailed balance condition (\ref{eq:db}). The only non-standard part of equation (\ref{eq:prob}) is the inclusion of the Jacobian~\cite{Borici}. Most applications use an area conserving molecular dynamics update, so that the logarithm of the Jacobian is zero. However, if it is possible to calculate the Jacobian, there is no restriction forcing the use of an area conserving algorithm, should an alternative method prove to be advantageous. 

\subsection{The new algorithm}
For simplicity, I start by considering a system with two eigenvectors of $Q$, $|\psi_1\rangle$ and $|\psi_2\rangle$, with eigenvalues $\lambda_1$ and $\lambda_2$. I intend to differentiate the eigenvector with respect to the gauge field, which requires finding the change in the eigenvectors caused by a small change in the gauge field. I write the new eigenvectors as
\begin{align}
|\psi_1'\rangle =&  |\psi_1\rangle \cos\theta + |\psi_2\rangle e^{i\delta}\sin\theta,\nonumber\\
|\psi_2'\rangle =&  |\psi_2\rangle \cos\theta - |\psi_1\rangle e^{-i\delta}\sin\theta.\label{eq:simplenotation}
\end{align}
If $\delta Q$ is the change in the kernel operator $Q$, and $\delta\lambda$ the change in the eigenvalue, then by considering the eigenvalue equations, 
\begin{align}
Q\ket{\psi_i} =& \lambda_i\ket{\psi_i},\nonumber\\
(Q+\delta Q)\ket{\psi'_i} = &(\lambda_i+\delta\lambda_i)\ket{\psi'_i},
\end{align}
it is easy to show that
\begin{align}
\tan2\theta =& \frac{2\sqrt{\langle\psi_2|\delta Q|\psi_1\rangle \langle\psi_1|\delta Q|\psi_2\rangle}}{\lambda_1-\lambda_2 + \langle\psi_1|\delta Q|\psi_1\rangle - \langle\psi_2|\delta Q|\psi_2\rangle}\label{eq:mixing}\\
\intertext{and}
e^{i\delta}=& \sqrt{\frac{\langle\psi_2|\delta Q|\psi_1\rangle}{\langle\psi_1|\delta Q|\psi_2\rangle}}.\label{eq:expand_this}
\end{align}
Using the usual equation of motion ($d/d\tau U = i \tau\Pi U$), it is possible to expand $\delta Q$ in $\tau$, which gives  
\begin{align}
\delta Q_{xy} =& -i\tau\kappa\gamma_5 \sum_{\mu}\big[(1-\gamma_\mu) \Pi_{\mu}(x) U_{\mu}(x)\delta_{y,x+\mu} -\nonumber\\
&\phantom{lots and lots of space} (1+\gamma_\mu)  U^{\dagger}_{\mu}(x-\mu)\Pi_{\mu}(x-\mu)\delta_{y,x-\mu}\big].
\end{align} 
The molecular dynamics momentum, $\Pi$, can be written as
\begin{gather}
\Pi_{\mu}(x) = \pi^{ix\mu}T^i_{\mu}(x),
\end{gather}
where $T^i_{\mu}(x)$ is a generator of $SU(3)$ (normalised so that $\text{Tr}\; T^iT^j = \delta_{ij}$) on a link proceeding from lattice site $x$ in direction $\mu$, and $\pi^{ix\mu }$ is a vector representation of the momentum field.  Now it is straightforward to express the change in the sign function in terms of the mixing angles $\theta$ and $\delta$:
\begin{align}
F^{NAC,\epsilon}_{ij}&\alpha_{ij}(\tau,\Pi)\pi\nonumber\\
=&|\psi_1'\rangle\langle\psi_1'|\epsilon(\lambda_1') + |\psi_2'\rangle\langle\psi_2'|\epsilon(\lambda_2') - (|\psi_1\rangle\langle\psi_1|\epsilon(\lambda_1) + |\psi_2\rangle\langle\psi_2|\epsilon(\lambda_2))\nonumber\\
=&|\psi_1\rangle\langle\psi_1|\sin^2\theta (\epsilon(\lambda_2) - \epsilon(\lambda_1))\frac{1}{2}\left(\frac{\langle\psi_1|\delta Q|\psi_2\rangle}{\langle\psi_1|\delta Q|\psi_2\rangle} + \frac{\langle\psi_2|\delta Q|\psi_1\rangle}{\langle\psi_2|\delta Q|\psi_1\rangle}\right) + \nonumber\\
&|\psi_2\rangle\langle\psi_2|\sin^2\theta (\epsilon(\lambda_1) - \epsilon(\lambda_2))\frac{1}{2}\left(\frac{\langle\psi_1|\delta Q|\psi_2\rangle}{\langle\psi_1|\delta Q|\psi_2\rangle} + \frac{\langle\psi_2|\delta Q|\psi_1\rangle}{\langle\psi_2|\delta Q|\psi_1\rangle}\right)+\nonumber\\
&|\psi_1\rangle\langle \psi_2| \cos\theta\sin\theta e^{-i\delta}(\epsilon(\lambda_1) - \epsilon(\lambda_2))\frac{\langle\psi_1|\delta Q|\psi_2\rangle}{\langle\psi_1|\delta Q|\psi_2\rangle} + \nonumber\\
&|\psi_2\rangle\langle \psi_1| \cos\theta\sin\theta e^{i\delta}(\epsilon(\lambda_1) - \epsilon(\lambda_2))\frac{\langle\psi_2|\delta Q|\psi_1\rangle}{\langle\psi_2|\delta Q|\psi_1\rangle}. \label{eq:expand_this2}
\end{align}  
$F^{NAC,\epsilon}_{ij}$, defined by this equation, shall be used to construct the NAC (non-area conserving) fermionic force. $\alpha_{ij}$ are defined below. The terms such as $\langle\psi_2|\delta Q|\psi_1\rangle/\langle\psi_2|\delta Q|\psi_1\rangle=1$ have been added for reasons that will be outlined in the discussion following equation (\ref{eq:21}). Note that by expanding equation (\ref{eq:expand_this2}) around $\tau = 0$, and neglecting terms of order $\tau^2$ and higher, one recovers the original expression for the derivative of the sign function (equation (\ref{eq:1})). But when $\theta$ becomes large, giving a large mixing between the two eigenvectors, this expansion breaks down. In order to construct a fermionic force from equation (\ref{eq:expand_this2}), I require the momentum vectors,
\begin{align}
\alpha_{ij}^{nx\mu} = &-i\kappa\tau\langle\psi_i|_x\gamma_5\big[(1-\gamma_\mu)T^n_{\mu}(x) U_{\mu}(x)\delta_{y,x+\mu} - \nonumber\\
&\phantom{lots and lots of space}(1+\gamma_{\mu}) U^{\dagger}_{\mu}(x)T^n_{\mu}(x)\delta_{x,y+\mu}\big]|\psi_j\rangle_y,
\end{align}
which are defined so that
\begin{gather}
\pi^{nx\mu}\alpha_{ij}^{nx\mu} = \bra{\psi_i}\delta Q\ket{\psi_j}.
\end{gather}
The (non-area conserving) fermionic force for this two eigenvalue system is thus 
\begin{gather}
F^{NAC}_{\mu}(x)(\tau,\Pi) = \langle X|\left[\gamma_5 F^{NAC,\epsilon}_{ij}(\tau,\Pi) + F^{NAC,\epsilon}_{ij}(\tau,\Pi) \gamma_5  | X\right]\rangle \alpha_{ij}^{nx\mu} T^{n}_{\mu}(x),
\end{gather} 
where $|X\rangle$ is the inverse of the overlap operator acting on the pseudo-fermion field $|\phi\rangle$:
\begin{gather}
|X\rangle = \frac{1}{H^2}|\phi\rangle,\label{eq:21}
\end{gather}
and the fermionic force $F_{\mu}(x)$ is defined as the quantity added to the old momentum to obtain the new momentum, i.e. $\Pi'_{\mu}(x) = \Pi_{\mu}(x) + F_{\mu}(x)$. $F^{NAC}$ refers to the term in $F$ which is constructed from the eigenvectors close to zero and not area conserving. The force is usually constructed as $F=-i\tau U\partial/\partial U (S_g + \phi H^{-2}\phi) + h.c.$, although in practice any Hermitian traceless matrix field which conserves energy and (up to a small, calculable, Jacobian) the measure will suffice. The dependence on the molecular dynamics time, $\tau$, is, in this notation, absorbed into the definition of the force. 

Now the reason why the terms equal to one have been added in equation (\ref{eq:expand_this2}) should be clear. We need a construction of the force such that $F^{NAC,\epsilon}_{ij}\pi\alpha_{ij}$ is indeed proportional to the momentum vectors, so that we can easily extract $F^{NAC,\epsilon}_{ij}$. This requires that the right hand side of  (\ref{eq:expand_this2}) is proportional to $\bra{\psi_i}\delta Q\ket{\psi_j}$ for some $i$ and $j$. Hence we have to introduce the additional terms equal to unity. The numerator of these terms provides the $\alpha_{ik}\pi_{ik}$ of the left hand side of (\ref{eq:expand_this2}). In principle, there is a choice between using \\
$\bra{\psi_1}\delta Q\ket{\psi_1}/\bra{\psi_1}\delta Q\ket{\psi_1}$ and $\bra{\psi_1}\delta Q\ket{\psi_2}/\bra{\psi_1}\delta Q\ket{\psi_2}$. However, we cannot introduce terms such as $\bra{\psi_1}\delta Q\ket{\psi_1}$ in the denominator of the force because of instabilities when this quantity becomes zero. Since $\sin^2\theta$ is proportional to $\bra{\psi_1}\delta Q\ket{\psi_2}$, there are no infinities in the definition of $F^{NAC,\epsilon}_{ij}$ given in equation (\ref{eq:expand_this2}), although obviously care is needed in its numerical implementation to avoid dividing zero by zero.

Of course, in real life we have more than two eigenvectors. Only the eigenvectors whose eigenvalues are close to zero need to be treated with the NAC algorithm. However, all eigenvectors with eigenvalues below a suitable cutoff, $\Lambda$, which has to be tuned for each set of simulation parameters, must be differentiated in this way. To include additional eigenvectors in the NAC setup, we need to include additional mixing angles. To simplify the expressions, I assume that only one mixing angle is large at any time, so that I can write the new eigenvector as
\begin{align}
\ket{\psi'_i} =&  \left(1+ \sum_{j\neq i}(\cos\theta_{ij}-1 )\right)\ket{\psi_i} + \sum_{j\neq i}\sin\theta_{ij}e^{i\delta_{ij}}\ket{\psi_j} +\phantom{a}\nonumber\\& \phantom{someextraspace}\frac{1}{Q-\lambda_i}\left(1-\sum_{j}\ket{\psi_j}\bra{\psi_j}\right)\delta Q \ket{\psi_i},\label{eq:psii}
\end{align} 
where the sum runs over all eigenvectors with eigenvalues below the cutoff. If there is more than one large mixing angle the new eigenvector defined in equation (\ref{eq:psii}) is no longer normalised. Although this problem has not occurred in my tests, the solution would be to use the full expansion in terms of Euler angles. For example, for three eigenvectors we would write 
\begin{gather}
\ket{\psi'_1} = \cos\theta_{12}\cos\theta_{13}\ket{\psi_1}+\cos\theta_{13}\sin\theta_{12}e^{i\delta_{12}}\ket{\psi_2} + \sin\theta_{13}e^{i\delta_{13}}\ket{\psi_3}.\label{eq:euler}
\end{gather}
For equation (\ref{eq:psii}), the mixing angles calculated in equations (\ref{eq:mixing}) and (\ref{eq:expand_this}) can be used. If equation (\ref{eq:euler}) is used, it would be necessary to derive new expressions for the mixing angles.

The non area-conserving force is a function of the momenta and is not an odd function  of the time. Therefore, to ensure reversibility it is necessary to update the momentum field in two steps:
\begin{align}
\Pi^{0.5} &= \Pi^0 +  F^{AC}\left(\frac{\tau}{2}\right) + F^ {NAC}\left(\frac{\tau}{2},\Pi^{0.5}\right),\nonumber\\
\Pi^{0.5} &= \Pi^1 +  F^{AC}\left(-\frac{\tau}{2}\right) + F^ {NAC}\left(-\frac{\tau}{2},\Pi^{0.5}\right).\label{eq:19}
\end{align}
The first step requires an iterative procedure. This iteration does not significantly slow down the HMC algorithm because the time-consuming parts of the force calculation, including the overlap inversions, eigenvalue calculation and the calculation of the momentum vectors $\alpha_{ij}$, are the same for each iteration and thus only need to be computed once for each calculation of the force. The iteration always converged to numerical precision within three or four steps. Given that the force is a highly non-linear function of the momentum, there is a danger that there may be multiple solutions to the iteration or chaotic effects. For this reason, the reversibility must be carefully checked. My numerical results on $8^316$ lattices are given in section \ref{sec:3.1}, and show no breakdown in reversibility across a large range of molecular dynamics time-steps.

Because this momentum update is not area conserving, two Jacobians must be calculated, one for each of the updates in equation (\ref{eq:19}). Both Jacobians can be computed using the same method. Since only the momentum is updated, $\partial U'/\partial U = 1$ and $\partial U'/\partial \Pi = 0$. Therefore, to calculate the Jacobian we need to calculate only $\partial \Pi'/\partial \Pi$. For the second update in (\ref{eq:19}), this is
\begin{align}
\frac{\partial\pi_{1}^{ix\mu }}{\partial\pi_{0.5}^{jy\nu }} =& \delta^{ix\mu ,jy\nu }+\alpha^{ix\mu}_{nm}\langle X|\left[\gamma_5 \frac{\partial}{\partial\pi_{0.5}^{jy\nu }}F^{NAC,\epsilon}_{nm}(-\tau/2,\Pi^{0.5}) +\phantom{a}\right. \nonumber\\&\phantom{lots of really really lovely space}\left.\frac{\partial}{\partial\pi_{0.5}^{jy\nu }}F^{NAC,\epsilon}_{nm}(-\tau/2,\Pi^{0.5}) \gamma_5 \right] | X\rangle\nonumber\\
=&\delta^{ij}\delta^{\mu\nu}\delta^{xy}  -\alpha^{ix\mu }_{nm}\alpha^{jy\nu }_{op} A_{nm,op}.
\end{align}
I obtain the second equality by noting that the only momentum dependence within $F$ is contained in terms such as $\bra{\psi_i} \delta Q\ket{\psi_j}$, which, when differentiated, gives terms proportional to $\alpha_{ij}$. By rewriting the vectors $\alpha_{ij}$ in terms of a complete orthonormal basis $\alpha'_{k}$, so that $\alpha_{ij}\alpha_{nm}A_{ij,nm} = \alpha'_k{\alpha'_l}^{\dagger}A'_{kl}$, it is easy to calculate the Jacobian in terms of the small matrix $A'$:
\begin{gather}
J = \det[1-A'].
\end{gather}
For sufficiently large eigenvalues, the logarithm of the Jacobian  should scale as O($\tau^3$) for each molecular dynamics step. This is the same as the change in the energy. The easiest way to see this is to note that the molecular dynamics update is reversible, which means that the logarithm of the Jacobian must be an odd function of time. Furthermore, at O($\tau$) this method is identical to the old area conserving algorithm; therefore the highest order term which can contribute to the Jacobian is O($\tau^3$). This is seen numerically in section \ref{sec:3.2}.

Of course, if the eigenvalues are small, and the Taylor expansion of $\sin\theta$ in $\tau/(\lambda_1-\lambda_2)$ does not converge, then it is possible to get large Jacobians, just as large forces blighted the old method. However, this method offers several advantages. Firstly, the change in the logarithm of the Jacobian scales as O($\log(\tau/(\lambda_1-\lambda_2))$), rather than a fermionic force (and thus change in kinetic energy) scaling as  $O(\tau/(\lambda_1-\lambda_2))$. Secondly the absence of large fermionic forces improves the stability of the algorithm (a small numerical error in a large force could lead to a large error in the energy). Finally, because the trajectory is smooth, there is a possibility of cancellations between a large positive Jacobian as the eigenvalues approach and a negative Jacobian as they depart; while with the old method the large force focused on one eigenvector meant that that eigenvector changed rapidly, leaving no opportunity for any cancellation. In our numerical tests on $8^316$ lattices I did not see any logarithms of Jacobians larger than 0.4 even at relatively large time steps. Energy violations of order 100 or higher were common with the old algorithm. These results will be discussed in section \ref{sec:3.3}.

In this paper, I have presented the method without any smearing, and it is not my intention to describe the smeared version of the algorithm in detail. However it is prudent to make a few comments. I have adapted and successfully run a version of this algorithm including stout smearing. From ~\cite{Morningstar:2003gk}, equation (71), I obtained an expression deriving the differential of the smeared link with respect to the differential of the original link. To calculate the vectors $\alpha^{n\mu x}_{ij}$, I simply applied this expression to the derivative of the gauge field, $i T^i U$. Equation (72) of ~\cite{Morningstar:2003gk}, which is normally used to calculate the smeared force, cannot be used because we need to efficiently calculate the Jacobian. While this approach can almost certainly be improved, it worked. Efficiently parallelising the code required adapting the algorithm so that it could calculate the differential of links separated by sufficient distance (twice the number of smearing steps plus one link) simultaneously. This procedure is acceptably quick for one or two smearing steps, but is more costly for larger iterations of smearing. 

\section{Numerical results}\label{sec:3}
I tested the algorithm on a $8^316$ ensemble with mass $\mu = 0.05$, $\beta = 8.35$ with a tadpole improved L\"uscher-Weisz gauge action~\cite{TILW,TILW2,TILW3,TILW4}, $\kappa = 0.2$, and no additional pseudo-fermions. In order to test the routine in the most extreme conditions possible on these lattices I did not use any stout smearing. In an actual HMC simulation, I would, of course, use moderate smearing to remove dislocations.

I will test the reversibility of the algorithm (section \ref{sec:3.1}), whether the Jacobian is sufficiently small to leave the acceptance rate unaffected, whether it scales with the molecular dynamics time as predicted (section \ref{sec:3.2}), and whether the new algorithm is indeed successful in eliminating the large forces (section \ref{sec:3.3}).
\subsection{Test of reversibility}\label{sec:3.1}
\begin{figure}
\begin{center}
\begin{tabular}{c}
\includegraphics[width=10cm]{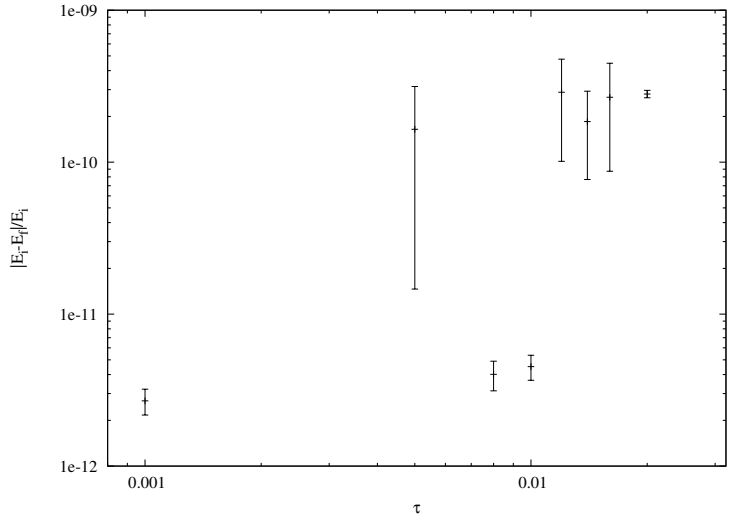}
\end{tabular}
\end{center}
\caption{Test of the reversibility. The plot shows the difference between the initial energy and the energy after running a forwards and backwards trajectory, normalised by the initial energy.}\label{fig:rev}
\end{figure}

To test that the algorithm is reversible, I ran forward and backward trajectories of length ten micro-canonical steps for twenty $8^316$ $\mu = 0.05$ configurations, and calculated the difference between the initial and final energies. I tested time steps between $\delta\tau = 0.001$ and $0.03$, and the average difference in the initial and final energies are plotted in figure \ref{fig:rev}. I see no breakdown in reversibility at any of these timescales (the energy differences are consistent with the accuracy which I use when inverting the overlap operator). I have also checked the reversibility by comparing the smallest Wilson eigenvalues during the forward and reverse trajectories; and again, there was no sign of a breakdown of reversibility to the working precision. 
\subsection{Scaling of Jacobian}\label{sec:3.2}
\begin{figure}
\begin{center}
\begin{tabular}{c}
\includegraphics[width=10cm]{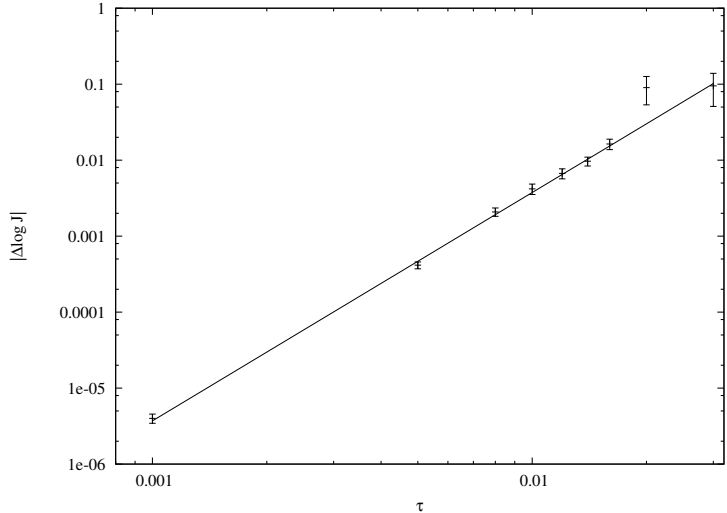}
\end{tabular}
\end{center}
\caption{The average change in the logarithm of the Jacobian for each micro-canonical step as a function of molecular dynamics time.}\label{fig:jac}
\end{figure}

To confirm that the Jacobian scales as expected with the molecular dynamics time, on the same configurations used in section \ref{sec:3.1}, I calculated the average change in the logarithm of the Jacobian, $\Delta\log J$, for each micro-canonical step. This average change is plotted against $\tau$ in figure $\ref{fig:jac}$, with the values given in table \ref{tab:jac}. To confirm that the scaling is the expected O($\tau^3$), I fitted the results using $|\Delta \log J| = (a \tau)^{n}$, using $a$ and $n$ as free parameters. The best fit, with seven degrees of freedom, had a $\chi^2$ value of 5.7. It gave $n = 3.005\pm 0.100$, the expected value within the statistical errors.  
\begin{table}
\begin{center}
\begin{tabular}{l l l l}
\hline
$\tau$&$\langle |\Delta \log J|\rangle$&$\langle \Delta \log J\rangle$ &$\max |\Delta \log J|$\\
\hline
0.001&$4.0(6)\times10^{-6}$&$1.5(9)\times 10^{-6}$&$3.14\times10^{-5}$\\
0.005&$4.1(4)\time10^{-4}$&$0.9(52)\times10^{-5}$&$3.11\times10^{-3}$\\
0.008&$2.1(3)\times10^{-3}$&$-0.2(39)\times10^{-4}$&0.02517\\
0.01&$4.2(7)\times10^{-3}$&$-4.0(76)\times 10^{-4}$&0.0636\\
0.012&$6.7(9)\times10^{-3}$&$-1.8(10)\times10^{-3}$&0.0818\\
0.014&$9.7(15)\times10^{-3}$&$-2.0(16)\times 10^{-3}$&0.0956\\
0.016&0.016(2)&$-5.8(23)\times10^{-3}$&0.139\\
0.02&0.090(36)&-0.018(39)&0.187\\
0.03&0.095(44)&0.013(51)&0.343\\
\hline
\end{tabular}
\end{center}
\caption{The average change in the absolute value of the logarithm of the Jacobian and the logarithm of the Jacobian for each micro-canonical step as a function of the molecular dynamics time, and the largest change in the Jacobian seen across the test trajectories on one micro-canonical step.}\label{tab:jac}
\end{table}
The largest change in the logarithm of the Jacobian for a micro-canonical step observed during the various test trajectories was 0.34: not large enough to cause the configuration to be rejected. The logarithm of the Jacobian, as can be seen in table \ref{tab:jac}, is not noticeably biased towards being either positive or negative. This means that over the course of a trajectory there will be cancellations between positive and negative $\log J$, so that the effect on the acceptance rate will be even smaller than might be expected from the O($\tau^3$) scaling.
\subsection{Comparison of fermionic forces}\label{sec:3.3}
\begin{figure}
\begin{center}
\begin{tabular}{c}
\includegraphics[height = 6.5cm]{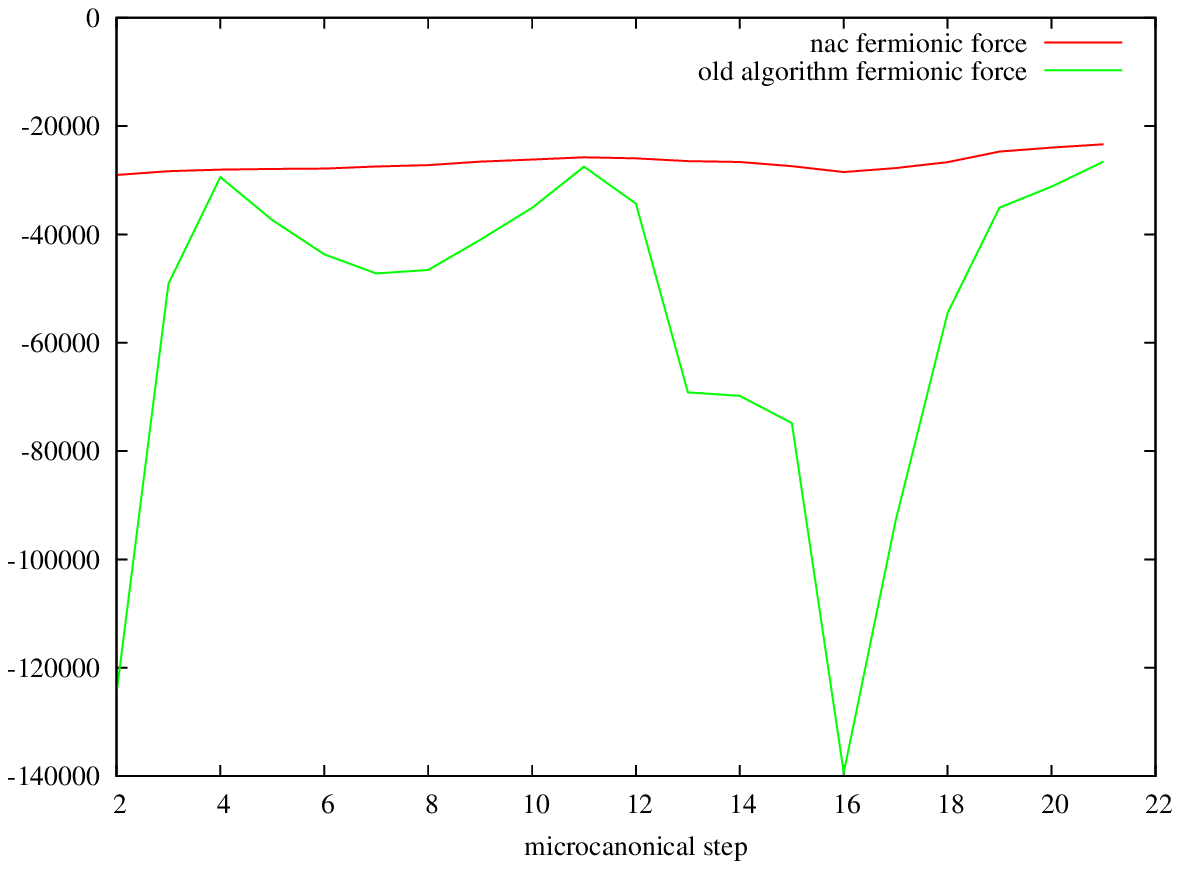}\\
\includegraphics[height = 6.5cm]{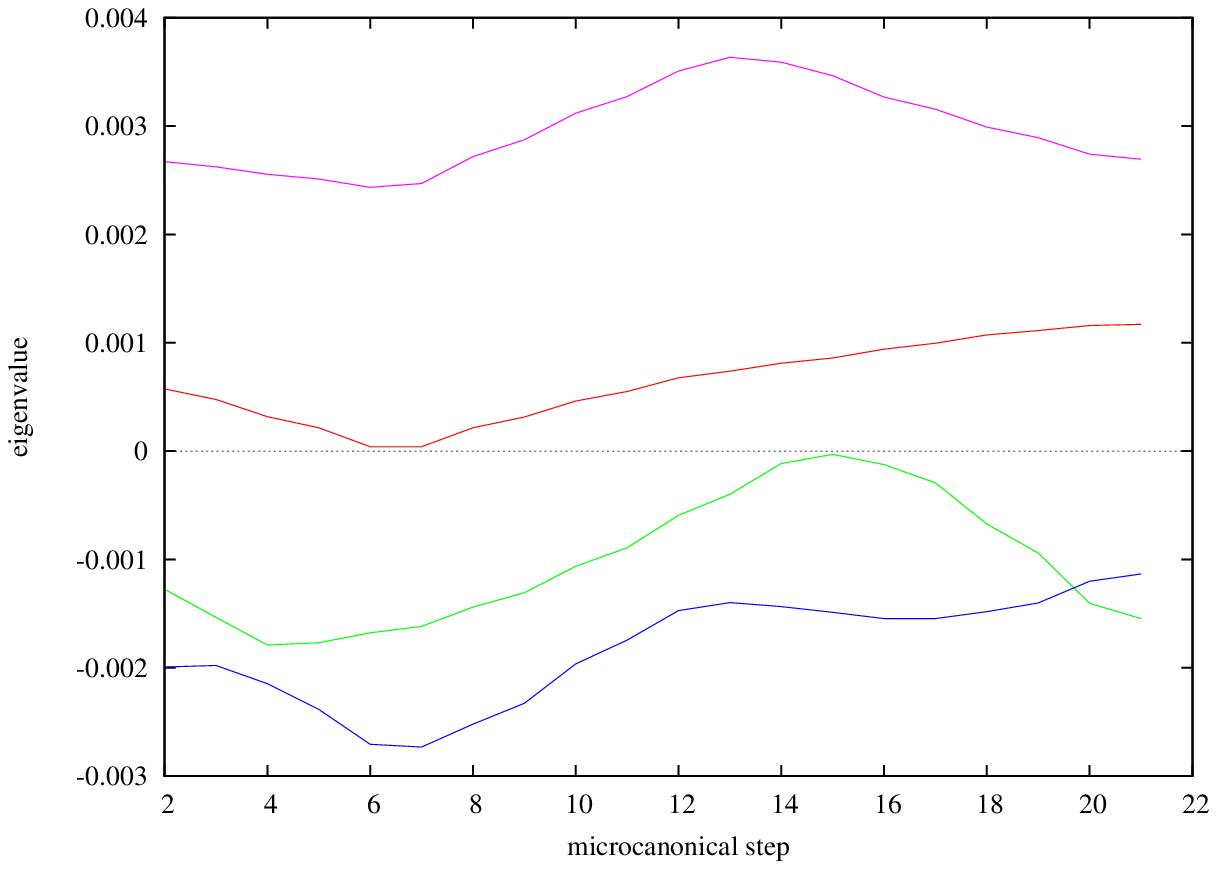}\\
\includegraphics[height = 6.5cm]{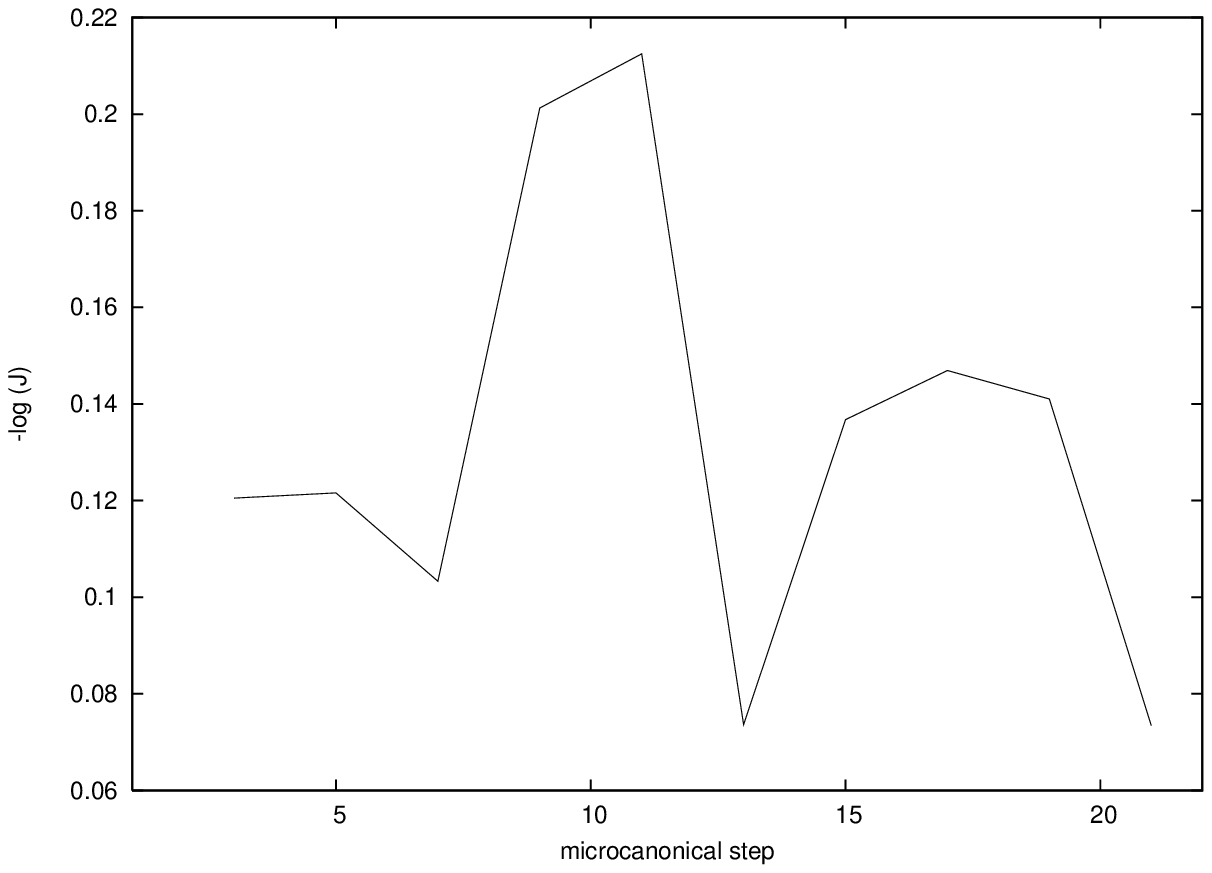}
\end{tabular}
\end{center}
\caption{Comparison of the trace of the square of the fermionic forces for the proposed and old algorithms with $\tau = 0.016$ on one of the $\mu = 0.05$  trajectories (top), together with the Wilson operator eigenvalues (middle) and the log of the Jacobian (bottom).}\label{fig:comparison}
\end{figure}

During my test trajectories, I calculated the fermionic force using both the original algorithm and the new algorithm, although I only used the force from the new algorithm when updating the momentum. This allowed me to directly compare the two forces. From figure \ref{fig:comparison} it is clear that the new fermionic force is stable, while the force from the old algorithm is considerably more unstable. The instabilities in the old algorithm fermionic force are, of course, exaggerated compared to a production run because I am not using any smearing (note that the eigenvalue scale in figure \ref{fig:oldforce}, based on data taken from a production run which used two levels of stout smearing, is a factor of ten larger than the scale in figure \ref{fig:comparison}). However, I expect the picture from figure \ref{fig:comparison} to be duplicated on larger lattices with smearing, because the density of smaller eigenvalues would increase. None of my test trajectories had any peaks in the fermionic force. As mentioned earlier, and as can be seen from the bottom plot in figure \ref{fig:comparison}, I did see peaks in the Jacobians caused by the mixing (as expected), but these were not large enough to reduce the metropolis acceptance rate.
\section{Conclusion}\label{sec:4}
I have presented a new method to differentiate the eigenvectors of the kernel operator in an Hybrid Monte Carlo algorithm with overlap fermions. This new algorithm is reversible, scales well with the molecular dynamics time, is no slower to compute than the old algorithm (unless an excessive number of smearing steps are used), and, unlike the old algorithm, has no large peaks in the fermionic force. The method can easily be extended to variants of the HMC algorithm, such as RHMC, using multiple pseudo-fermion, or differentiable smearing. I therefore recommend that this new method is used in future dynamical overlap calculations which allow small kernel eigenvalues. 
\subsection*{Acknowledgements}
I would like to thank Thomas Lippert, Andreas Sch\"afer, Stefan Krieg, Anna Hasenfratz , Tony Kennedy, Tom DeGrand, Stefan Schafer and especially Artan Borici for many useful discussions. I would also like to thank the referee for his many helpful comments on and corrections to the first versions of this paper. The numerical calculations were carried out on the cray-XD1 at the John von Neumann institute for computing at the Forschungszentrum J\"ulich. I was supported by grant 930183 from the EU RP-6 ``Hadron Physics" project, from the DFG ``Gitter-Hadronen Ph\"anomenologie" project, number 458/14-4 and EU grant MC-EIF-CT-2004-506078 during the course of this research.
\appendix
\section{Calculation of force and Jacobian}\label{sec:5}
In this appendix, for simplicity I concentrate on the force and Jacobian from the mixing of one eigenvector pair. However, to illuminate the generalisation to the multiple eigenvector case of equation(\ref{eq:psii}), and to avoid confusion between $\theta_{ij}$ and $\theta_{ji}$, I maintain the notation of equation (\ref{eq:psii}) rather than reverting to the simpler notation of equation (\ref{eq:simplenotation}). The argument outlined here can easily be extended to include other pairs of eigenvectors. I also only consider the momentum update from $\Pi^{0.5}$ to $\Pi^1$, since the fermionic force and Jacobian  for the update from $\Pi^0$ to $\Pi^{0.5}$ can be constructed in the same way.

I write the force in terms of the momentum vectors
\begin{align}
\alpha_{ij}^{nx\mu} =& -i\kappa\tau\langle\psi_i|_x\gamma_5\big[(1-\gamma_\mu)T^n_{\mu}(x) U_{\mu}(x)\delta_{y,x+\mu} -\nonumber\\
&\phantom{lots and lots of space here} (1+\gamma_{\mu}) U^{\dagger}_{\mu}(x)T^n_{\mu}(x)\big]\delta_{y,x-\mu}|\psi_j\rangle_y,
\end{align}
where $T^n$ are the Gell-Mann matrices normalised so that $\text{Tr}(T^n T^m) = \delta_{nm}$.

Neglecting the gauge action and area conserving fermionic action, the energy conservation equation for an update from fields $[\Pi,U]$ to $[\Pi',U']$ reads
\begin{align}
0=&\frac{1}{2}({\Pi'}^2 - \Pi^2) +\bra{\phi}\frac{1}{{H[U']}^2}\ket{\phi} - \bra{\phi}\frac{1}{{H[U]}^2}\ket{\phi}\\ 
\approx&\pi^{nx\mu}\left[({\pi'}^{nx\mu} - \pi^{nx\mu}) - F_{ii}^{nx\mu} - F_{jj}^{nx\mu} - F_{ij}^{nx\mu} - F_{ji}^{nx\mu}\right],
\end{align}
where 
\begin{align}
\pi^{nx\mu} =& \text{Tr}(T^n \Pi_{\mu}(x)),& \Pi_{\mu}(x) =& T^n \pi^{nx\mu}\label{eq:pipi}
\end{align}
 and
\begin{align}
F_{ii}^{nx\mu} = &\frac{1}{2}C_{ii}\sin^2\theta_{ij}\left(\frac{\alpha_{ji}^{nx\mu}}{\delta Q_{ji}} + \frac{\alpha_{ij}^{nx\mu}}{\delta Q_{ij}}\right),\nonumber
\\
F_{ij}^{nx\mu} = &C_{ij}\sin\theta_{ij}\cos\theta_{ij} e^{-i\delta_{ij}} \frac{\alpha_{ij}^{nx\mu}}{\delta Q_{ij}},\nonumber
\\
F_{ji}^{nx\mu} = &C_{ji}\sin\theta_{ij}\cos\theta_{ij} e^{i\delta_{ij}}\frac{\alpha_{ji}^{nx\mu}}{\delta Q_{ji}},\nonumber
\\
F_{jj}^{nx\mu} = &-\frac{1}{2}C_{jj}\sin^2\theta_{ij}\left(\frac{\alpha^{nx\mu}_{ji}}{\delta Q_{ji}} + \frac{\alpha_{ij}^{nx\mu}}{\delta Q_{ij}}\right),\nonumber\\
C_{ab}=&(1-\mu^2)\left(\bra{X}\gamma_5\ket{\psi_a}\braket{\psi_b}{X} + \braket{X}{\psi_a}\bra{\psi_b}\gamma_5\ket{X}\right)\left(\epsilon(\lambda_i) - \epsilon(\lambda_j)\right),\nonumber\\
\delta Q_{ab} = & \bra{\psi_a}\delta Q\ket{\psi_b}
.\label{eq:thefs}
\end{align}
%Need to explain how I get this.
Equation (\ref{eq:pipi}) can be used to convert between the vector form of the momentum (more useful in this formulation) and the matrix form (used in the numerical implementation). $\delta Q$, $\theta$ and $\delta$ are all functions of $\Pi$.

Energy is conserved if
\begin{align}
{\pi'}^{nx\mu} =& \pi^{nx\mu} + F_{ii}^{nx\mu} + F_{ij}^{nx\mu} + F_{ji}^{nx\mu} + F_{jj}^{nx\mu}\\
=&\pi^{nx\mu}+B_{ij}\alpha_{ij}^{nx\mu},
\end{align}
where the coefficients $B_{ij}$ can be determined from equation (\ref{eq:thefs}). 

To calculate the Jacobian, $\partial \pi^{nx\mu}/\partial\pi^{my\nu}$ I note that
\begin{align}
\frac{\partial \delta Q_{ii}}{\partial\pi^{my\nu}} =& \alpha_{ii}^{my\nu},\nonumber\\
\frac{\partial \delta Q_{ji}}{\partial\pi^{my\nu}} =& \alpha_{ji}^{my\nu},\nonumber\\
\frac{\partial \delta Q_{ij}}{\partial\pi^{my\nu}} =& \alpha_{ij}^{my\nu},\nonumber\\
\frac{\partial \delta Q_{jj}}{\partial\pi^{my\nu}} =& \alpha_{jj}^{my\nu},\\
\intertext{giving}
\frac{4}{\sin4\theta_{ij}}\frac{\partial\theta_{ij}}{\partial\pi^{my\nu}} =& \frac{\alpha_{ji}^{my\nu}}{2\delta Q_{ji}} + \frac{\alpha_{ij}^{my\nu}}{2\delta Q_{ij}} -\frac{\alpha_{ii}^{my\nu} - \alpha_{jj}^{my\nu}}{\lambda_i - \lambda_j + \delta Q_{ii} - \delta Q_{jj}}\\
\intertext{and}
e^{-i\delta_{ij}}\frac{\partial e^{i\delta_{ij}}}{\partial\pi^{my\nu}}=&\frac{\alpha_{ji}^{my\nu}}{2\delta Q_{ji}} - \frac{\alpha_{ij}^{my\nu}}{2\delta Q_{ij}}.
\end{align}
I use these expressions to differentiate $B_{ij}$, and write the Jacobian in the form
\begin{gather}
\frac{\partial {\pi'}^{nx\mu}}{\partial\pi^{my\nu}} = \delta_{n,m}\delta_{xy}\delta_{\mu\nu} - \alpha^{nx\mu}_{ij}\alpha^{my\nu}_{op}A_{ij,op},
\end{gather}
where
\begin{align}
A_{ji,ii} = &\frac{\sin4\theta_{ij}\left((C_{jj}-C_{ii})\sin2\theta_{ij} - 2C_{ji}\cos2\theta_{ij} e^{i\delta_{ij}}\right)}{8\delta Q_{ji}(\lambda_i-\lambda_j + \delta Q_{ii}-\delta Q_{jj})},
\nonumber\\
A_{ji,ji} = &\frac{(C_{jj} - C_{ii})\left(8\sin^2\theta_{ij} - \sin2\theta_{ij}\sin4\theta_{ij}\right)}{16(\delta Q_{ji})^2}   -
\phantom{a}\nonumber\\&\phantom{lotsandlotsofspace}
 \frac{C_{ji}e^{i\delta_{ij}}(2\sin2\theta_{ij} - \cos2\theta_{ij}\sin4\theta_{ij}) }{8(\delta Q_{ji})^2},
\nonumber\\
A_{ji,ij} = &\frac{(C_{jj}-C_{ii})\sin2\theta_{ij}\sin4\theta_{ij}}{16(\delta Q_{ji})(\delta Q_{ij})} + \frac{C_{ji}e^{i\delta_{ij}}(\cos2\theta_{ij}\sin4\theta_{ij} - 2\sin2\theta_{ij}) }{8(\delta Q_{ji})(\delta Q_{ij})} ,
\nonumber\\
A_{ji,jj} = &-A_{ji,ii},
\nonumber\\
A_{ij,ii} = &\frac{\sin4\theta_{ij}\left((C_{jj}-C_{ii})\sin2\theta_{ij} - 2C_{ij}\cos2\theta_{ij} e^{-i\delta_{ij}}\right)}{8\delta Q_{ij}(\lambda_i-\lambda_j + \delta Q_{ii}-\delta Q_{jj})},
\nonumber\\
A_{ij,ji} = &\frac{(C_{ii}-C_{jj})\sin2\theta_{ij}\sin4\theta_{ij}}{16(\delta Q_{ji})(\delta Q_{ij})} + \frac{C_{ij}e^{-i\delta_{ij}}\left(\cos2\theta_{ij}\sin4\theta_{ij} - 2 \sin2\theta _{ij} \right)}{8(\delta Q_{ji})(\delta Q_{ij})},
\nonumber\\
A_{ij,ij} = &\frac{(C_{jj} - C_{ii})\left(8\sin^2\theta_{ij} -\sin2\theta_{ij}\sin4\theta_{ij} \right)}{16(\delta Q_{ij})^2} -
 \phantom{a}\nonumber\\&\phantom{lotsandlotsofspace}
\frac{C_{ij}e^{-i\delta_{ij}}\left( 2 \sin2\theta_{ij}  - \cos2\theta_{ij}\sin4\theta_{ij}\right)}{8(\delta Q_{ij})^2},
\nonumber\\
A_{ij,jj} = &-A_{ij,ii}
,
\end{align}
and all other components of $A$ are 0. 

It is my intention to use the standard result
\begin{gather}
\left|1-\alpha^{nx\mu}_i{\alpha_j^{my\nu}}^{\dagger}A_{ij}\right| = |1-A_{ij}|\label{eq:standard_result}
\end{gather}
to calculate the determinant. There are two things which must be done before applying  this result. First of all, I have calculated $\alpha_{ij}\alpha_{nm}A_{ij,nm}$ not $\alpha_i\alpha^{\dagger}_jA_{ij}$; however since  $\alpha_{ij} = \alpha_{ji}^{\dagger}$, the correct expression is obtained by exchanging the $12$ and $21$ columns of the matrix $A_{ij,nm}$ calculated above. Secondly, I need to re-express the $\alpha$s in terms of an orthonormal basis. It is easy, though only necessary in the theoretical proof of equation (\ref{eq:standard_result}) and not in a numerical implementation, to construct other vectors orthonormal to the new $\alpha$s so that the basis spans the entire space.

I can construct an orthonormal basis for $\alpha_{ij}$ by, first of all, expressing the vectors in terms of a single index, and writing
\begin{align}
\alpha_1 \rightarrow &\alpha_1/\sqrt{(\alpha_1,\alpha_1)},\nonumber\\
A_{1i} \rightarrow & A_{1i}\sqrt{(\alpha_1,\alpha_1)},\nonumber\\
A_{i1}\rightarrow & A_{i1}\sqrt{(\alpha_1,\alpha_1)}, 
\end{align}
then projecting $\alpha_1$ from the other vectors $\alpha_j$
\begin{align}
x = & (\alpha_1,\alpha_j);\nonumber\\
\alpha_j \rightarrow& \alpha_j - x \alpha_1;\nonumber\\
A_{ii}\rightarrow &A_{ii} +x^{\dagger}A_{1j} + xA_{j1} + A_{jj}xx^{\dagger},
\nonumber\\
A_{i1}\rightarrow &A_{i1}+x^{\dagger}A_{ij},
\nonumber\\
A_{1i}\rightarrow &A_{1i}+xA_{ji},
\nonumber\\
A_{1j}\rightarrow &A_{1j}+xA_{jj},
\nonumber\\
A_{j1}\rightarrow &A_{j1}+x^{\dagger}A_{jj},
\end{align}
for $i\neq 1$ and $i\neq j$. This procedure can then be repeated for the other vectors in turn. Once recast into an orthonormal basis, I can use equation (\ref{eq:standard_result}) to express the Jacobian in terms of the determinant of a small matrix. This determinant can then be easily calculated using a standard method, for example LU decomposition~\cite{NumericalrecipiesLU}.
\section{The reflection/transmission update}
During the transmission step, which occurs when an eigenvalue of the kernel operator crosses zero (and the momentum is sufficiently large that I do not have to reflect), I recommend using an momentum update
\begin{align}
\Pi^+ =& \Pi^- + \tau_c(F) - \eta \tau_c(\eta,F)+ \nonumber\\
&  \left(\eta_1 (\eta_1, \Pi^- + \frac{\tau_c}{2}(F^+ + F^-)) + \eta_2 (\eta_2, \Pi^-
+ \frac{\tau_c}{2}(F^+ + F^-))\right)\times\nonumber\\
&\left[\sqrt{1 +
  \frac{d_2 }{(\eta_1,
    \Pi^- + \frac{\tau_c}{2}(F^+ + F^-))^2 +  (\eta_2, \Pi^- +
    \frac{\tau_c}{2}(F^++F^-)^2}}-1\right]+\nonumber\\
    &\eta(\Pi^-,\eta)\left[\sqrt{-2\frac{\log\left(e^{-(\Pi^-,\eta)^2/2 - 2d}+1-e^{-2d} \right)}{(\Pi^-,\eta)^2}}- 1\right],
%\nonumber\\
%&-\left(\eta_1^2 (\eta_1^2, \Pi^- - \frac{\tau_c}{2}(F^- + F^+)) +\eta_2^2 (\eta_2^2, \Pi^-
%- \frac{\tau_c}{2}(F^- + F^+))\right)
\label{eq:40}\\
d_2 =& ( - 2\tau_c(F^-,\eta)(\Pi^-,\eta)
  +2\tau_c(F^+,\eta)(\Pi^+,\eta) + \tau_c^2(F^- + F^+,F)).
\end{align}
The notation, which is chosen to be consistent with my earlier work, is outlined in~\cite{Cundy:2005pi,Cundy:2005mr} together with the full details of the construction and why I believe it to be superior to other algorithms. Here I will limit myself to explaining the most important features of each term.
 In this formula (in the case where there is no smearing, and up to a normalisation factor) $\eta=\alpha_{ii}$ is a unit momentum vector normal to the surface of zero eigenvalue (in the space of all possible gauge field configurations), $d$ is half of the change to the momentum energy, $\eta_1$ and $\eta_2$ are arbitrary vectors perpendicular to $\eta$ and the force difference $F=(F^+-F^-) - \frac{1}{3}\text{Tr}(F^+-F^-)$, the $-$ superscript indicates a force or momentum calculated with the smallest eigenvalue having its original sign, while $+$ indicates that the force or momentum was calculated with the eigenvalue having its final sign, and all these quantities are calculated on the gauge field with zero eigenvalue. In this section, when referring to $\alpha_{ij}$ in general (for example in the calculation of the Jacobians), it should be understood that I am excluding $\eta = \alpha_{ii}$.  Even with the old algorithm, this update is not area conserving, but it is constructed to conserve the action, including the Jacobian term. The $d_2$ term cancels out an O($\tau$) energy difference caused because the momentum is not updated at the moment of crossing~\cite{Cundy:2005pi}. The other improvement to the algorithm originally published by Zoltan Fodor \textit{et al.}~\cite{Fodor:2003bh} is in the term proportional to $\eta$, which has an increased rate of transmission~\cite{Cundy:2005mr}. 

However, this update is a function of the fermionic force, and by using the non area conserving fermionic force proposed in this paper, it is necessary to calculate the Jacobian for the transmission step. For simplicity, I shall here write $F$ as a function of $\Pi^-$, although in practice, to maintain reversibility, it is again necessary to update the momentum in two steps, using an iterative procedure for one of the updates. First of all, I need to construct an orthonormal basis, $\tilde{\alpha}_{ij}$, $\eta_1$ and $\eta_2$ from the vectors $\alpha_{ij}$, $\alpha_{ii}$ and two additional vectors, where I ensure that $\eta_1$ and $\eta_2$ are both also normal to the area conserving part of $F$ (the non-area conserving part of $F$ is of course proportional to the vectors $\alpha_{ij}$ in any case). For convenience, I write that $\tilde{\alpha}_{ii} = \alpha_{ii} = \eta$. It is easy to show that because $\eta_1$ and $\eta_2$ are normal to $F$ and all the vectors $\alpha_{ij}$, $\frac{\partial(\Pi^+,\tilde{\alpha}_{ij})}{\partial(\Pi^-,\eta_k)} = 0$. Similarly, $\frac{\partial(\Pi^+,\eta)}{\partial(\Pi^-,\alpha_{ij})} = 0$, (except, of course, when $i = j = 1$) and $\frac{\partial(\Pi^+,\eta)}{\partial(\Pi^-,\eta_k)} = 0$. Thus, I write the Jacobian in the form
\begin{align}
J = \left|\begin{array}{c c c}
\frac{\partial (\Pi^+,\eta)}{\partial (\Pi^{-},\eta)}&\frac{\partial (\Pi^+,\eta)}{\partial (\Pi^{-},\tilde{\alpha}_{ij})}&\frac{\partial (\Pi^+,\eta)}{\partial (\Pi^{-},\eta_k)}\\
\frac{\partial (\Pi^+,\tilde{\alpha}_{ij})}{\partial (\Pi^{-},\eta)}&\frac{\partial (\Pi^+,\tilde{\alpha}_{ij})}{\partial (\Pi^{-},\tilde{\alpha}_{ij})}&\frac{\partial (\Pi^+,\tilde{\alpha}_{ij})}{\partial (\Pi^{-},\eta_k)}\\
\frac{\partial (\Pi^+,\eta_k)}{\partial (\Pi^{-},\eta)}&\frac{\partial (\Pi^+,\eta_k)}{\partial (\Pi^{-},\tilde{\alpha}_{ij})}&\frac{\partial (\Pi^+,\eta_k)}{\partial (\Pi^{-},\eta_k)}
\end{array}\right| = 
\left|\begin{array}{c c c}
\frac{\partial (\Pi^+,\eta)}{\partial (\Pi^{-},\eta)}&0&0\\
\neq 0&\frac{\partial (\Pi^+,\tilde{\alpha}_{ij})}{\partial (\Pi^{-},\tilde{\alpha}_{ij})}&0\\
\neq 0&\neq 0&\frac{\partial (\Pi^+,\eta_k)}{\partial (\Pi^{-},\eta_k)}
\end{array}\right|.
\end{align}   
${\partial (\Pi^+,\eta)}/{\partial (\Pi^-,\eta)}$ and ${\partial (\Pi^+,\eta_i)}/{\partial (\Pi^-,\eta_i)}$ have already been calculated in ~\cite{Cundy:2005pi,Cundy:2005mr}. All that remains is to calculate the Jacobian for ${\partial (\Pi^+,\tilde{\alpha}_{ij})}/{\partial (\Pi^+,\tilde{\alpha}_{ij})}$. For one of these two half-updates, I obtain
\begin{align}
(\Pi^+,\tilde{\alpha}_{ij})=&(\Pi^{0.5},\tilde{\alpha}_{ij})+\tau_c(F(\Pi^{0.5}),\tilde{\alpha}_{ij})\nonumber\\
=&(\Pi^{0.5},\tilde{\alpha}_{ij})+\tau_c((F(\Pi^{0.5}),\alpha_{k})- (F(\Pi^{0.5}),\eta)(\eta,\alpha_{k})).
\end{align}
Thus,
\begin{gather}
\frac{\partial(\Pi^+,\tilde{\alpha}_{k})}{\partial(\Pi^+,\tilde{\alpha}_{n})}=\delta_{k,n}+\frac{1}{2}A'_{k,n}\alpha'_{k}{\alpha'}^{\dagger}_{n}-\frac{1}{2}A'_{kn}(\alpha'_n,\eta)(\eta,\alpha'_k).
\end{gather}
And from here, I proceed as before.

\bibliographystyle{modified_cpc}
\bibliography{nacpaper}

\end{document}